# A modified approach to the measurement problem: Objective reduction in the retinal molecule *prior* to conformational change

Fred H. Thaheld

*fthaheld@directcon.net*

## Abstract

A new analysis of the measurement problem reveals the possibility that collapse of the wavefunction may now take place just *before* photoisomerization of the rhodopsin molecule in the retinal rods. It is known that when a photon is initially absorbed by the retinal molecule which, along with opsin comprises the rhodopsin molecule, an electron in the highest $\pi$ orbital is immediately excited to a $\pi^*$ orbital. This means that a measurement or transfer of information takes place at the quantum level *before* the retinal molecule commences the conformational change from *cis* to *trans*. This could have profound implications for resolving some of the foundational issues confronting quantum mechanics.

## Introduction

As some of you may recall, for several years the author and others have been advocates of a collapse or measurement process taking place within the rhodopsin molecule or the retinal rod cells of the eye, so that only objective information is ever presented to the brain, mind or consciousness (Adler, 2006; Shimony, 1998; Thaheld, 2005, 2006, 2007).  This means that none of these entities are required to collapse the wavefunction.

To briefly recapitulate from prior papers, the photopigment in rod cells is called rhodopsin, which is composed of two components (Kandel et al, 2000; Thaheld, 2005, 2007).  The first is a membrane bound protein molecule called opsin, which is covalently bound to a second component called 11-*cis* retinal, which is a derivative of vitamin A (Wald, 1968).  The 11-*cis* retinal molecule has 6 alternating single and double bonds making up a long unsaturated electron network (Kandel et al, 2000).  A rhodopsin molecule has a molecular weight of about $4 \times 10^4$ nucleons and a diameter of about $4 \times 10^{-7}$ cm (Adler, 2006).  The 11-*cis* retinal chromophore (light harvesting molecule) consists of some 40 atoms and has an active site of ~10 Angstroms (Mathies, 1999, 2004; Thaheld, 2005).  Let us now introduce a photon into this picture and see what happens.

**The new approach**

A photon (one can also use the plural) that has been emitted or scattered by the text projected on a computer screen or printed on a sheet of paper, carries information of this text at the quantum or microscopic level (Zurek, 2007).  This photon could have also acquired information from an infinite number of sources in the universe.  When this photon is absorbed by the retinal molecule into one of the $\pi$ bonds found between the 11$^{th}$ and 12$^{th}$ carbon atoms, an electron in the highest $\pi$ orbital is immediately excited from a $\pi$ to a higher energy $\pi^*$ orbital.  Several interesting things will have happened as a result of the absorption of this photon *prior* to the conformational change:

The former superposed state no longer exists as a result of the collapse of the wavefunction; we are still at the quantum level; the conservation of energy has not been violated; we can identify the site where the collapse has taken place; and this change is irreversible! One now has to ask, before proceeding any further, as to whether this can be considered a measurement at the quantum level, just a simple transference of information or both? And, if so, just what constitutes an 'apparatus' at the quantum level or is such a term, now even needed (Bell, 1990)?

Since the electron is now excited to an antibonding $\pi^*$ orbital, the carbons between C=11 and C=12 are now able to rotate freely around this bond. This allows retinal to change its conformation from *cis* to *trans* in ~200 fs (Mathies, 1999), with the result that this all-*trans* retinal configuration which was formerly "bent", is now straightened and does not fit into the binding site of the opsin molecule (Chang, 1998). As a result upon isomerization, which is a transformation of a molecule into a different isomer or molecular arrangement, the *trans* isomer separates from the opsin molecule and a series of changes in the protein begins. As the opsin protein molecule changes its conformation, it initiates a cascade of biochemical reactions that result in the closing of $Na^+$ channels in the rod cell membrane (Baylor, 1996). Prior to this event $Na^+$ ions flow freely into the cell to compensate for the lower potential (more negative charge) which exists inside the rod cell. When the $Na^+$ channels are closed, a large potential difference builds up, with the inside of the cell becoming more negative as the outside of the cell becomes more positive (Whikehart, 2003). This potential difference is passed along as an electrical impulse. Thus it is that one photon activated rhodopsin molecule causes approximately $10^6$ charges or sodium ions to fail to enter the rod cell, resulting in an amplified electrical current about 1 pA in amplitude lasting ~200 ms (Rieke, Baylor, 1998). This amounts to $2 \times 10^{-13}$ C, with $1.6 \times 10^{-19}$ C (which is the elementary unit of charge) per sodium ion (Rieke, 2008). This means that the information has been amplified from the microscopic to the mesoscopic level and, that at some point in the amplification process, depending upon the number of ions, crosses over into the macroscopic or classical world. This response results in between 2-3 action potentials in the optic nerve, all carrying this amplified information. We will be able to *approximately* determine the *cuts,* or what may be referred to as the *seamless transition* between the microscopic-mesoscopic and mesoscopic-macroscopic worlds.

Matsuno has commented upon this theory and has also said something similar in previous papers (Matsuno, 1989, 1996, 2007). "The collapse of the wavefunction is an extremely elementary and ubiquitous phenomenon in the quantum regime. Any transition of an orbital electron moving around an atom or a molecule through absorption of a photon, is an instance of a collapse of the photon wavefunction. An example is the photoelectric effect. Even if the wave front of the photon wavefunction is far greater than the atomic scale, the photon has to shrink its own body down to the atomic size when the metallic plate irradiated by the photon flux emits electrons. There is no need to invoke CSL, GRW or Everett's splitting to get the collapse of the wavefunction right. What is relevant to biology is the process of *amplification of the collapse* once initiated elsewhere. This has been stated in a similar fashion in terms of the inseparability of the equation of motion and its boundary conditions (Matsuno, 1989, 1996). Preparation of boundary conditions always comes with the collapse of the wavefunction when viewed from the perspective of the unitary dynamics applied to quantum phenomena."

**Conclusion**

1. It would appear that neither decoherence, the environment, gravity (Salart et al, 2008) CSL, GRW or Bohm has any role to play in this *animate* collapse process, from the quantum beginning initiated by photoexcitation, to the macroscopic ending proposed in this theory (Matsuno, 1989, 1996, 2007; Thaheld, 2006). And, since reduction does take place in this fashion, this would rule out the Everett relative state or many worlds theory (Thaheld, 2005).

2. That due to the small size of the 11-*cis* retinal chromophore with its 40 atoms and an active site of ~10 Angstroms, the collapse of the wavefunction takes place at the quantum level, involving one of the atoms or molecules. The proof of this collapse is revealed when a π orbital electron is excited to a π* orbital. However, can one still use the term 'apparatus' to describe either these atoms or molecules, since we are not at the classical level (Bell, 1990; Thaheld, 2007)?

3. It appears that there are three *cuts* that take place that would be measurable. The first *cut* representing the actual collapse taking place between the photon and the atom or molecule, leading to a π* orbital electron. The second *cut* taking place between the quantum and mesoscopic worlds at ~$10^{-7}$ m or ~$10^2$ nm (the size of a virus). And the third *cut* taking place between the mesoscopic and macroscopic worlds at ~$10^{-6}$ m or ~$10^3$ nm (the size of a bacterium). The latter two *cuts* occur during the amplification process, with *irreversibility* occuring at all three *cuts*. Since there is a grey area involving the exact boundaries where the latter two *cuts* might take place, as the size scale of interest has no rigid definition, it may be more appropriate to use the term *seamless transition* instead, especially when one is dealing with *amplification* following the collapse of the wavefunction.

4. Since all animal eyes share a common molecular strategy using opsin for catching photons and, opsins appeared in biological systems *before* eyes, the theory advanced in this paper should also be applicable to *any living entity* with or without eyes, vertebrate or invertebrate (Fernald, 2004; Thaheld, 2005). The author has previously addressed this issue with regards to the *Euglena gracilis*, a unicellular protozoan dating back about 2 billion years (Thaheld, 2005;Wolken, 1967). It possesses two different photoreceptors, an eyespot or stigma for light searching, and chloroplast for photosynthesis. This analysis could also be extended to and include the evolution of primitive 'eyes' of cyanobacteria, which are photosynthetic, some 3.5 billion years ago (Gehring, 2001). This probably means that the signal transduction cascade or methods of amplification relied upon by such primitive organisms after collapse of the wavefunction, would be of a much simpler and/or quicker nature.

5. And, it is important to stress once again, that it is the *amplification of the collapse* once it is initiated elsewhere, that is relevant to biology (Matsuno, 2007). It is this *amplification* which brings this information from the quantum level to the classical world. So it is that one arrives at an interdisciplinary answer to the measurement problem involving physics, chemistry and biology.

## Acknowledgements

I am indebted to the reviewers and the editor for their comments and suggestions which helped to improve the quality of this paper. To Thesa von Hohenastenberg-Wigandt whose tenacity in the face of adversity served as an extra inducement in the author's quest. Finally, to the lady B.K., who reminded me that "all things are possible".

**References**

Adler, S., 2006. Lower and upper bounds on CSL parameters from latent image formation and IGM heating. quant-ph/0605072.

Baylor, D., 1996. How photons start vision. Proc. Acad. Sci. U.S.A. 93, 560-565.

Bell, J.A., 1990. Against 'measurement'. Phys. World. Aug. 33-40.

Chang, R., 1998. Chemistry, 6$^{th}$ Ed. McGraw Hill, N.Y.

Fernald, R.D., 2004. Evolving eyes. Int. J. Dev. Biol. 48, 701-705.

Gehring, W.J., 2001. The genetic control of eye development and its implications for the evolution of the various eye-types. Zoology 104, 171-183.

E.R. Kandel, J.H. Schwartz, T.M. Jessell, 2000. *Principles of Neural Science*, 4$^{th}$ edn. McGraw-Hill, New York. 507-522. (See especially p. 511, Fig. 26-3 and p. 515, Fig. 26-6).

Mathies, R.A., 1999. Photons, femtoseconds and dipolar interactions: A molecular picture of the primary events in vision. *Rhodopsin and Phototransduction*. John Wiley and Sons, N.Y.

Mathies, R.A., 2003, 2004. Private communication.

Matsuno, K., 1989. *Protobiology: Physical Basis of Biology.* Boca Raton: CRC Press (See especially Chap. 2 – Internal Measurement, 31-55).

Matsuno, K., 1996. Internalist stance and the physics of information. BioSystems 38, 111-118.

Matsuno, K., 2007. Private communication.

Rieke, F. and Baylor, D., 1998. Single-photon detection by rod cells of the retina. Rev. Mod. Phys. 70, 1027-1036.

Rieke, F., 2008. Private communication.

Salart, D., Baas, A., van Houwelingen, J.A.W., Gisin, N., Zbinden, H., 2008. Space-like separation in a Bell test assessing gravitationally induced collapses. quant-ph/0803.2425.

Shimony, A., 1998. Comments on Leggett's "Macroscopic Realism", in: *Quantum Measurement: Beyond Paradox.* R.A. Healey and G. Hellman, eds. Univ. of Minnesota, Minneapolis. p. 29.

Thaheld, F.H., 2005. Does consciousness really collapse the wavefunction? A possible objective biophysical resolution of the measurement problem. BioSystems 81, 113. quant-ph/0509042.

Thaheld, F.H., 2006. The argument for an objective wavefunction collapse: Why spontaneous localization collapse or no-collapse decoherence cannot solve the measurement problem in a subjective fashion. quant-ph/0604181.

Thaheld, F.H., 2007. Nature as the observer: A simplified approach to the measurement problem. quant-ph/0702114.

Wald, G., 1968. The molecular basis of visual excitation. Nature 219, 800-807.

Whikehart, D., 2003. *Biochemistry of the Eye*, 2nd ed. Butterworth-Heineman, Philadelphia.

Wolken, J., 1967. *Euglena: An Experimental Organism for Biochemical and Biophysical Studies*. 2nd ed. Meredith Publishing Co., N.Y.

Zurek, W.H., 2007. Quantum origin of quantum jumps: Breaking of unitary symmetry induced by information transfer and the transition from quantum to classical. quant-ph/0703160.